\documentclass[12pt]{iopart}
\usepackage{amssymb,mathptm,dcolumn,bm,times,graphicx,color}

\begin{document}

\title{Microwave response of superconducting pnictides: extended $s_{\pm}$
scenario}
\author{O.V.~Dolgov$^{1}$}
\author{A.A.~Golubov$^{2}$}
\author{D.~Parker$^{3}$}
\address{$^1$Max-Planck-Institut f\"{u}r Festk\"{o}rperforschung, D-70569 Stuttgart,
Germany}
\address{$^2$Faculty of Science and Technology and MESA+ Institute of Nanotechnology,
University of Twente, 7500 AE Enschede, The Netherlands}
\address{$^3$Naval Research Laboratory, 4555 Overlook Ave. SW, Washington, DC 20375}

\begin{abstract}
We consider a two-band superconductor with relative phase $\pi $ between the two order
parameters as a model for the superconducting state in ferropnictides.
Within this model we calculate the microwave response and the NMR relaxation rate.
The influence of intra- and interband impurity scattering beyond the
Born and unitary limits is taken into account.
We show that, depending on the scattering rate, various types of power law
temperature dependencies of the magnetic field penetration depth and the NMR relaxation rate
at low temperatures may take place.
\end{abstract}

\maketitle

\section{Introduction}

The recent discovery of Fe-based superconducting compounds \cite{kamihara}
has stimulated the research of unconventional superconductors. One of the
most important and still unsettled issues is the symmetry of the superconducting
gap function. So far, different experiments produce conflicting results. As
regards measurements of the penetration depth and the NMR relaxation rate, a
power law behavior at low temperatures is now clearly established, which is
a signature of unconventional order parameter symmetry. One possible
scenario of a pairing symmetry state is a superconductor consisting of two
relatively small semimetallic Fermi surfaces, separated by a finite wave
vector $\mathbf{Q}$ with the relative phase $\pi $ between the two order
parameters. This is the so-called $s_{\pm }$ model, first proposed in Ref.~%
\cite{mazin}. In our previous work \cite{Parker} we have shown that $s_{\pm
} $ model with strong impurity scattering can explain the power law behavior
of the NMR relaxation rate. Therefore it is important to extend this formalism
to address microwave properties of a two-band $s_{\pm }$ superconductor, in
particular the magnetic field penetration depth and real part of complex
conductivity, since experimental data are now available for
single crystals of Fe-based superconductors.

In this paper we calculate the microwave response and the NMR relaxation rate
for a model $s_{\pm }$ superconductor in which impurity scattering is
treated beyond the Born limit and discuss the relevance to the experimental data
for Fe-based superconducting compounds.

\section{General expressions}

We describe a multiband superconductor in the framework of the Eliashberg
approach equations for the renormalization function $Z_{i}(\omega )$ and
complex order parameter $\phi _{i}(\omega ).$   As shown in the first reference
of \cite{gaps}, the BCS approach
can give highly inaccurate results in the case of interband superconductivity due to the BCS neglect of mass renormalization.  In addition there is evidence for strong-coupling in the pnictides,
with many experimentally determined $\Delta/T_{c}$ ratios substantially exceeding the BCS value of 1.76,
and so we therefore employ the Eliashberg equations.

On the real frequency axis
they have the following form, assuming an \textit{uniform} ( band-independent) impurity
scattering (see e.g., Ref.\cite{Parker,PM,GolMaz})
\begin{eqnarray}
\phi _{i}(\omega ) &=&\sum_{j}\int\limits_{-\infty }^{\infty
}dzK_{ij}^{\Delta }(z,\omega )\mathrm{Re}g_{j}^{\Delta }(z)+\mathrm{i}\frac{%
\gamma }{2\mathcal{D}}\left( g_{1}^{\Delta }(\omega )-g_{2}^{\Delta }(\omega
)\right)  \nonumber \\
(Z_{i}(\omega )-1)\omega &=&\sum_{j}\int\limits_{-\infty }^{\infty
}dzK_{ij}^{Z}(z,\omega )\mathrm{Re}g_{j}^{Z}(z)+\mathrm{i}\frac{\gamma }{2%
\mathcal{D}}\left( g_{1}^{Z}(\omega )+g_{2}^{Z}(\omega )\right) ,  \label{ee}
\end{eqnarray}%
where $\mathcal{D}=1-\sigma +\sigma \left[ \left( g_{1}^{Z}(\omega
)+g_{2}^{Z}(\omega )\right) ^{2}+\left( g_{1}^{\Delta }(\omega
)-g_{2}^{\Delta }(\omega )\right) ^{2}\right] $. For our model $%
g_{i}^{Z}(\omega )=n_{i}\left( \omega \right) Z_{i}(\omega )\omega
/D_{i}(\omega )$, $g_{i}^{\Delta }(\omega )=n_{i}\left( \omega \right) \phi
_{i}(\omega )/D_{i}(\omega )$, where $D_{j}(\omega )=\sqrt{\left[
Z_{j}(\omega )\omega \right] ^{2}-\phi _{j}^{2}(\omega )}$ and $n_{i}\left(
\omega \right) $ is a partial density of states. $\gamma =2c\sigma/\pi N(0)$
is the normal-state scattering rate, N(0) is the total density of states
(i.e.summed over both bands) at the Fermi level, {\it c} is the impurity concentration,
and $\sigma =$ $\frac{\left[ \pi N(0)v\right] ^{2}}{1+\left[ \pi N(0)v\right]
^{2}}$ is the impurity strength ($\sigma \rightarrow 0$ corresponds to the
Born limit, while $\sigma =1$ to the unitary one). The kernels $K_{ij}^{\Delta
,Z}(z,\omega )$ describe the \textit{electron-boson interaction} and have forms

\[
K_{ij}^{\Delta ,Z}(z,\omega )=\int\limits_{0}^{\infty }d\Omega \frac{\tilde{B%
}_{ij}(\Omega )}{2}\left[ \frac{\tanh \frac{z}{2T}+\coth \frac{\Omega }{2T}}{%
z+\Omega -\omega -i\delta }-\left\{ \Omega \rightarrow -\Omega \right\} %
\right] ,
\]%
where the spin-fluctuation coupling function is $\tilde{B}_{ij}(\Omega )=B_{ij}(\Omega )=
=\lambda _{ij}{\pi \omega \Omega _{sf}}/({\Omega _{sf}^{2}+\omega
^{2}})$ for the equation for $\phi $%
, and $\left\vert B_{ij}(\Omega )\right\vert $ for the equation for $Z$. Here $\lambda_{ij}$ is the coupling constant pairing band $i$ with band $j$ and  $\Omega_{SF}$ is the spin fluctuation frequency.  Note
that all retarded interactions enter the equations for the renormalization
factor $Z$ with a positive sign.

We note that the implementation of the band-independent impurity scattering is contained in the second term on the right-hand side of Eq. 1, where the $\gamma$ is applied to both bands (albeit with
a relative minus sign in the first equation due to the order parameter sign change between bands).  We have chosen such a band-independent scattering for several reasons, including consistency with the previously published work and to avoid a proliferation of parameter choices.  However, recent work of Senga and Kontani \cite{senga} suggests that this assumption is justified on an experimental basis.  Their Fig. 4 shows that only $\gamma_{inter}/\gamma_{intra}$ between 0.9 and 1 is consistent with the several sets of nuclear spin relaxation rate T$_{1}^{_1}$ data showing $T^{2.5}-T^{3.0}$ behavior over a very large temperature range.  The theoretical rationale for such a comparatively large interband scattering rate remains unclear, but can be plausibly related to the inherent disorder in these systems, with the dopant atoms themselves acting as scattering centers.

\textbf{The microwave conductivity }in the London (local, $\mathbf{q}\equiv
0 $) limit is given by
\begin{equation}
\sigma ^{i}(\omega )=\omega _{pl,i}^{2}\Pi _{i}(\omega )/4\pi i\omega ,
\label{sig}
\end{equation}%
where $\Pi _{i}(\omega )$ is an analytical continuation to the real
frequency axis of the polarization operator (see, e.g. Refs. \cite{Nam},\cite%
{LRZ},\cite{DGS},\cite{AC},\cite{mars1})
\[
\Pi _{i}(\omega )=\left\{ i\pi T\sum_{n}\Pi _{i}(\omega _{n},\nu
_{m})\right\} _{i\omega _{m}\Longrightarrow \omega +i0^{+}},
\]%
\begin{eqnarray}
\Pi _{i}(\omega ) &=&\int d\omega ^{\prime }\left\{ \frac{\tanh \left( \frac{%
\omega _{-}}{2T}\right) }{D^{R}}\right\vert \left[ 1-\frac{\tilde{\omega}%
_{-}^{R}\tilde{\omega}_{+}^{R}{\large +}\phi _{-}^{R}\phi _{+}^{R}}{\sqrt{(%
\tilde{\omega}_{\_}^{R})^{2}\mathbf{-(}\phi _{-}^{R})^{2}}\sqrt{(\tilde{%
\omega}_{+}^{R})^{2}\mathbf{-(}\phi _{+}^{R})^{2}}}\right] -  \label{real} \\
&&\frac{\tanh \left( \frac{\omega _{+}}{2T}\right) }{D^{A}}\left[ 1-\frac{%
\tilde{\omega}_{-}^{A}\tilde{\omega}_{+}^{A}\mathbf{+}\phi _{-}^{A}\phi
_{+}^{A}}{\sqrt{(\tilde{\omega}_{\_}^{A})^{2}\mathbf{-(}\phi _{-}^{A})^{2}}%
\sqrt{(\tilde{\omega}_{+}^{A})^{2}\mathbf{-(}\phi _{+}^{A})^{2}}}\right] -
\nonumber \\
&&\frac{\tanh \left( \frac{\omega _{+}}{2T}\right) -\tanh \left( \frac{%
\omega _{-}}{2T}\right) }{D^{a}}\left\vert \left[ 1-\frac{\tilde{\omega}%
_{-}^{A}\tilde{\omega}_{+}^{R}\mathbf{+}\phi _{-}^{A}\phi _{+}^{R}}{\sqrt{(%
\tilde{\omega}_{\_}^{A})^{2}\mathbf{-(}\phi _{-}^{A})^{2}}\sqrt{(\tilde{%
\omega}^{R})^{2}\mathbf{-(}\phi _{+}^{R})^{2}}}\right] \right\} ,  \nonumber
\end{eqnarray}%
where
\[
D^{R,A}=\sqrt{(\tilde{\omega}_{+}^{R,A})^{2}\mathbf{-(}\phi _{+}^{R,A})^{2}}+%
\sqrt{(\tilde{\omega}_{\_}^{R,A})^{2}\mathbf{-(}\phi _{-}^{R,A})^{2}},
\]%
and
\[
D^{a}=\sqrt{(\tilde{\omega}_{+}^{R})^{2}\mathbf{-(}\phi _{+}^{R})^{2}}-\sqrt{%
(\tilde{\omega}_{\_}^{A})^{2}\mathbf{-(}\phi _{-}^{A})^{2}},
\]%
$\omega _{\pm }=\omega ^{\prime }\pm \omega /2$, and the index $R(A)$
corresponds to the retarded (advanced) brunch of the complex function $%
F^{R(A)}={Re}F\pm i{Im}F$ ( the band index $i$ is omitted), and $\tilde{\omega}
=Z_{i}(\omega)\omega$. Here $
\omega _{pl}^{\alpha \beta }=\sqrt{8\pi e^{2}\langle N_{i}(0)v_{F}^{\alpha
}v_{F}^{\beta }\rangle }$ is the plasma frequency in different directions.
For the dirty case the low frequency limits of
expressions \ref{sig} and \ref{real} can be reduced to the strong coupling generalization
of the famous Mattis-Bardeen expressions \cite{mattis}
\begin{eqnarray}
\sigma _{1}(\omega &\rightarrow &0)=\sigma _{1}^{\textrm{dc}%
}\int\limits_{0}^{\infty }d\omega \left( -\frac{\partial f(\omega )}{%
\partial \omega }\right) \left\{ \left[ \mathrm{Re}g_{1}^{Z}(\omega )\right]
^{2}+\left[ \mathrm{Re}g_{1}^{\Delta }(\omega )\right] ^{2}\right\} +
\nonumber \\
&&\sigma _{2}^{\textrm{dc}}\int\limits_{0}^{\infty }d\omega \left( -\frac{%
\partial f(\omega )}{\partial \omega }\right) \left\{ \left[ \mathrm{Re}%
g_{2}^{Z}(\omega )\right] ^{2}+\left[ \mathrm{Re}g_{2}^{\Delta }(\omega )%
\right] ^{2}\right\} ,
\label{sig_imp}
\end{eqnarray}%
where $\sigma _{i}^{\textrm{dc}}=N_{i}(0)v_{F}^{2}e^{2}\tau _{i}$ is a contribution to
the static conductivity from $i-$th band. Note that in the London limit
there are no \textit{cross-terms} connected two bands.

An important characteristic of the superconducting state is \textbf{the
penetration depth }of the magnetic field $\lambda _{L,\alpha \beta }$ in the
local (London) limit, which is related to the imaginary part of the optical
conductivity by
\begin{equation}
1/\lambda _{L,\alpha \beta }^{2}=\lim_{\omega ->0}4\pi \omega \mathop{\rm Im}%
\sigma ^{\alpha \beta }(\omega ,\mathbf{q}=0)/c^{2}\equiv \omega
_{pl,i}^{\alpha \beta 2}{Re}\Pi _{i}(\omega =0)/c^{2},
\label{eq:pen-london}
\end{equation}%
where $\alpha ,\beta $ denote again Cartesian coordinates and $c$ is the
velocity of light. If we neglect strong-coupling effects (or, more
generally, Fermi-liquid effects) then for a clean uniform superconductor at $%
T=0$ we have the relation $\lambda _{L,\alpha \beta }=c/\omega _{pl}^{\alpha
\beta }$. Impurities and interaction effects drastically enhance the
penetration depth, and it is suitable to introduce a so called 'superfluid
plasma frequency' $\omega _{pl,\alpha \beta }^{sf}$ by the relation $\omega
_{pl,\alpha \beta }^{sf}=c/\lambda _{L,\alpha \beta }$. It has been often
mentioned that this function corresponds to the charge density of the
superfluid condensate, but we would like to point out that this is only the
case for noninteracting clean systems at $T=0$.

In the two-band model we have the standard expression (neglecting vertex
corrections)
\begin{eqnarray}
1/\lambda _{L,\alpha \beta }^{2}(T) &\equiv &(\omega _{pl,\alpha \beta
}^{sf}(T)/c)^{2}=  \label{eq:pen} \\
&&\sum_{i}\left( \frac{\omega _{pl,i}^{\alpha \beta }}{c}\right) ^{2}\pi
T\sum_{n=-\infty }^{\infty }\frac{\tilde{\Delta}_{i}^{2}(n)}{[\tilde{\omega}%
_{i}^{2}(n)+\tilde{\Delta}_{i}^{2}(n)]^{3/2}},  \nonumber
\end{eqnarray}%
where $\tilde{\omega}(n)$ and $\tilde{\Delta}(n)$ are the solutions of Eq.\ %
\ref{ee} continued to the imaginary (Matsubara) frequencies ($\tilde{\Delta}%
_{i}(n)=\phi _{i}(i\omega _{n}),$ $\tilde{\omega}_{i}(n)=\omega
_{n}Z_{i}(i\omega _{n})$). The calculations along these formulas can be
thus presented in form of the effective superfluid plasma frequency, $\omega
_{pl}^{sf}.$

For \textbf{the NMR\ relaxation rate}, following \cite{SM}, we can write down the following general expressions.

\begin{equation}
1/T_{1}T=-\frac{1}{2\pi }\lim_{\omega \rightarrow 0}\sum_{\mathbf{q}}\left[
F(\mathbf{q})\right] ^{2}\frac{{Im}\chi _{\pm }(\mathbf{q},\omega )}{%
\omega },  \label{t1}
\end{equation}%
where $\chi _{\pm }(\mathbf{q},\omega )$ is an analytical continuation to
the real axis of the Fourier \ transform of the correlator%
\[
\chi _{\pm }(\mathbf{r},\tau )=-\left\langle \left\langle T_{\tau }(S_{+}(%
\mathbf{r},-i\tau )S_{-}(\mathbf{0},0)\right\rangle \right\rangle _{imp}.
\]%
averaged over the impurity ensemble. Here $S_{\pm }(\mathbf{r},-i\tau )=\exp
(H\tau )S_{\pm }(\mathbf{r})\exp (-H\tau )$ where $H$ is the electron
Hamiltonian, $\tau $ denotes imaginary time, and $S_{+}(\mathbf{r})=\psi
_{\uparrow }^{\dag }(\mathbf{r})\psi _{\downarrow }(\mathbf{r})$ and $S_{-}(%
\mathbf{r})=\psi _{\downarrow }^{\dag }(\mathbf{r})\psi _{\uparrow }(\mathbf{%
r}).$ As a result we have

\begin{eqnarray}
1/T_{1}T &=&\frac{1}{\pi ^{2}}\sum_{\mathbf{k}_{1},\mathbf{k}%
_{2}}\int_{-\infty }^{\infty }d\omega (-\frac{\partial f(\omega )}{\partial
\omega })\sum_{i,j}\left[ F_{ij}(\mathbf{k}_{1}-\mathbf{k}_{2})\right]
^{2}\times  \nonumber \\
&&\left[ {Im}\frac{\omega Z_{i,\mathbf{k}_{1}}(\omega )}{D_{i,\mathbf{k}%
_{1}}(\omega )}{Im}\frac{\omega Z_{j,\mathbf{k}_{2}}(\omega )}{D_{j,%
\mathbf{k}_{2}}(\omega )}\right. +  \nonumber \\
&&{Im}\frac{\xi _{i,\mathbf{k}_{1}}}{D_{i,\mathbf{k}_{1}}(\omega )}%
{Im}\frac{\xi _{j,\mathbf{k}_{2}}}{D_{j,\mathbf{k}_{2}}(\omega )}+
\label{t3} \\
&&\left. {Im}\frac{\phi _{i,\mathbf{k}_{1}}(\omega )}{D_{i,\mathbf{k}%
_{1}}(\omega )}{Im}\frac{\phi _{j,\mathbf{k}_{2}}(\omega )}{D_{j,%
\mathbf{k}_{2}}(\omega )}\right] .  \nonumber
\end{eqnarray}%
Here $D_{i,\mathbf{k}_{1}}(\omega )=\left[ \omega Z_{i,\mathbf{k}%
_{1}}(\omega )\right] ^{2}-\xi _{i,\mathbf{k}_{1}}^{2}-\phi _{i,\mathbf{k}%
_{1}}^{2}(\omega )$, $\xi _{i,\mathbf{k}_{1}}$is the bare energy. For the
Fermi-contact interaction
\begin{equation}
\frac{1}{T_{1}T}\propto \int\limits_{0}^{\infty }d\omega \left( -\frac{%
\partial f(\omega )}{\partial \omega }\right) \left\{ \left[ \mathrm{Re}%
g_{1}^{Z}(\omega )+\mathrm{Re}g_{2}^{Z}(\omega )\right] ^{2}+\left[ \mathrm{%
Re}g_{1}^{\Delta }(\omega )+\mathrm{Re}g_{2}^{\Delta }(\omega )\right]
^{2}\right\} .  \label{t4}
\end{equation}%
This expression
contains\ \textit{the cross-term} in contrast to the microwave conductivity.
In this paper, in the T$_{1}^{_1}$ calculation only these cross terms are used to emphasize the interband
character of the superconductivity, as it is these cross terms that are most enhanced by the
nearly antiferromagnetic state within a more detailed RPA approximation.
For a single band system the full expression is proportional to Eq.\ref{sig_imp} when $\sigma
_{1}^{\textrm{dc}}\rightarrow \infty $ (Ref. \cite{mars2}), but in multiband systems $1/T_{1}T$
and $\sigma _{1}(\omega \rightarrow 0)$ can behave differently.

\section{Results and discussion}

It is well known that pair-breaking impurity scattering can induce
substantial sub-gap density of states, which can produce power-law low
temperature behavior in a whole host of thermodynamic quantities, such as
specific heat, London penetration depth, nuclear spin relaxation rate, and
even optical conductivity. Such behavior has been well-studied in the two
canonical limits of weak (Born) scattering and strong (unitary) scattering
\cite{hirsh} , but the intermediate regime has received almost no attention.
In addition, with the advent of the multiband superconductivity in MgB$_{2}$
and the apparent multiband, \textit{primarily interband} superconductivity
in the pnictides, comes a need for further study of the intermediate regime
in an interband case. Recent studies \cite{Bang,VVC,Chubukov} have addressed the
effects of impurities in the pnictides, but only in the Born or unitary
limits. Here we study the important and likely more realistic intermediate
regime, with $\sigma $, effectively the scattering strength is varied in the range
from $\sigma = 0$ corresponding to the Born limit to $\sigma = 1$ corresponding to the
unitary limit.  As stated earlier, for all calculations
the impurity scattering rate $\gamma_{intra}=\gamma_{inter}=0.8\Delta_{0}$.

We will now illustrate the above discussion using specific numerical models.
First, we present numerical solutions of the Eliashberg equations using the
spin-fluctuation model for the spectral function of the intermediate boson: $%
B_{ij}(\omega )=\lambda _{ij}{\pi \omega \Omega _{sf}}/({\Omega _{sf}^{2}+\omega
^{2}})$, with the parameters $\Omega _{sf}=25$meV, $\lambda _{11}=\lambda
_{22}=0.5$, and $\lambda _{12}=\lambda _{21}=-2$.   The rather large coupling constants are
an attempt to model the rather large experimentally observed ratio $\Delta/T_{c}$.  This set gives a
reasonable value for $T_{c}\simeq 26.7$K. A similar model was used in Ref.~%
\cite{timusk} to describe optical properties of ferropnictides. This model was also
\begin{figure}[tbp]
\includegraphics[width=0.8\linewidth]{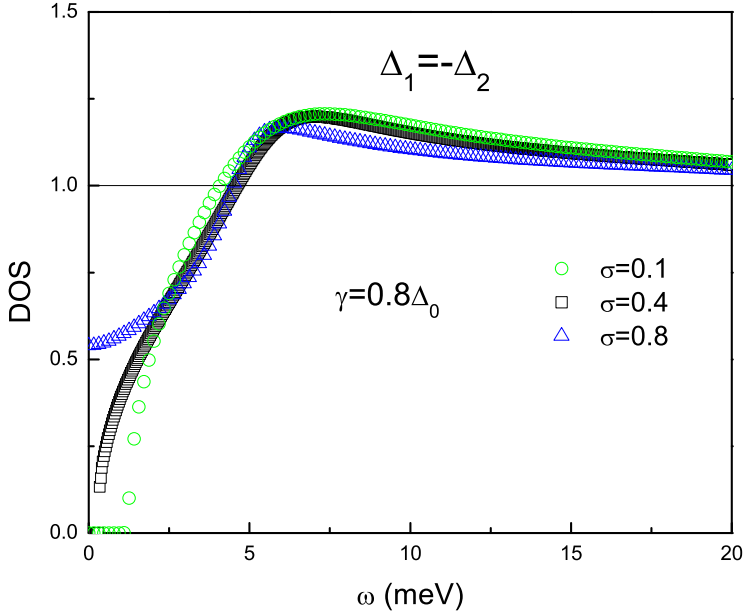}
\caption{(color online) The quasiparticle density of states for the three
indicated cases.  The near-Born case $\sigma=0.1$ retains a small gap, while the intermediate case shows a monotonic DOS and the near-unitary is gapless.}
\label{fig2}
\end{figure}
used in \cite{Parker} and for consistency is used here.
As stated earlier, we further assume that each surface features the same gap \cite{gaps}, and that the intraband impurity scattering rate and interband scattering rate are both equal to $0.8\Delta_{0}$, where $\Delta_{0}$ is the low-temperature limiting value of the superconducting gap $\Delta$.As in \cite{Parker}, we have chosen a relatively large impurity scattering, which is to be expected considering the early state
of pnictide sample preparation and the limited availability of large single
crystals.
\begin{figure}[tbp]
\includegraphics[width=0.8\linewidth]{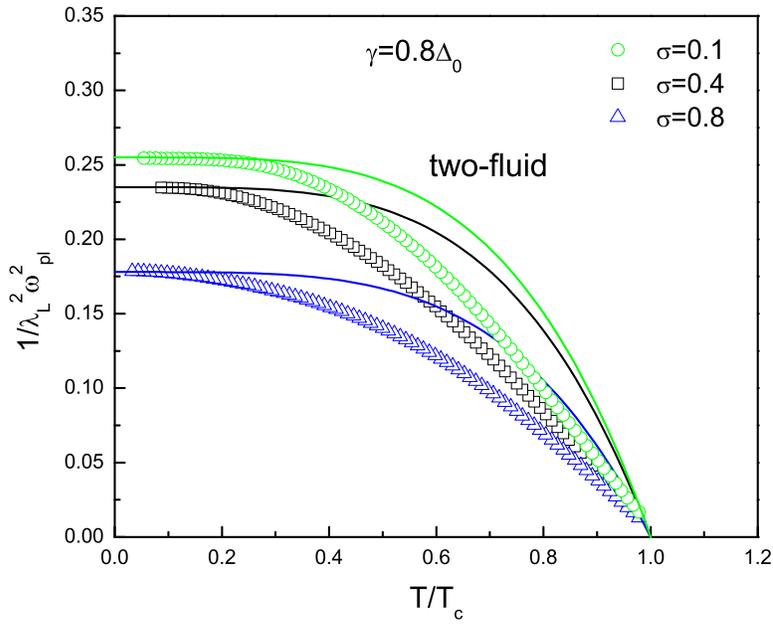}
\caption{(color online) The inverse squared penetration depth.  The near-Born limit approaches the BCS "two-fluid" calculation ($\propto 1-T^{4}$)  at low temperatures, mimicking exponential behavior, while the other two cases show power-law behavior, as in Fig. 3.}
\label{fig2}
\end{figure}
\begin{figure}[tbp]
\includegraphics[width=0.8\linewidth]{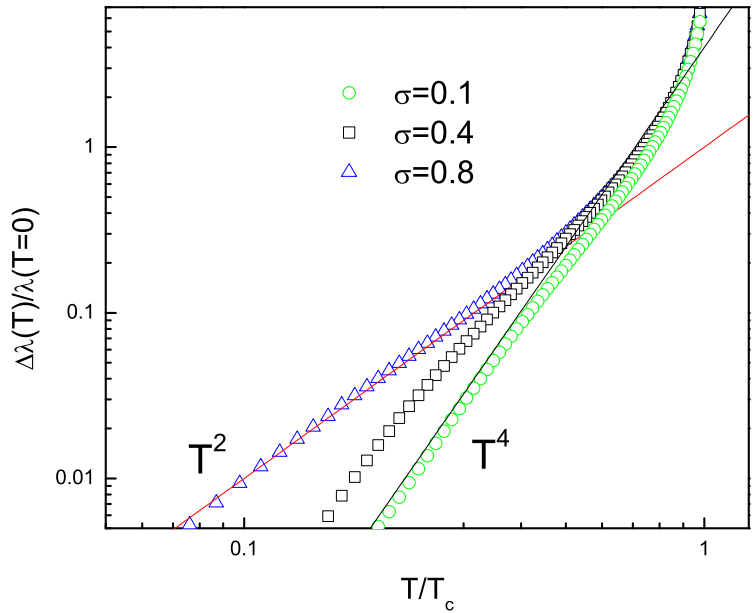}
\caption{(color online) The low temperature behavior of the penetration depth, showing the evolution with decreasing $\sigma$ from T$^{2}$ gapless behavior towards the exponential-mimicking T$^{4}$ character.}
\label{fig3}
\end{figure}

We begin with the density of states, shown below in Figure 1. Several effects are apparent. Firstly, for all three $\sigma$
values the substantial peak usually present at $\omega=\Delta_{0}$ (about 6
meV here) is substantially truncated, with much spectral weight transferred
below the gap. However, the detailed sub-gap behavior depends radically upon
the scattering strength $\sigma$. The near-Born case $\sigma=0.1$ still
retains a small minigap of approximately 1.5 meV, which will lead to
exponentially activated behavior below about 4 Kelvin. Although some data
has shown evidence for such exponentially activated behavior, there is also
significant data showing power-law behavior. The intermediate case $%
\sigma=0.4$ shows a monotonically increasing density of states and
essentially no minigap, leading to power-law behavior, as proposed in \cite%
{Parker}. Finally, the near-unitary case $\sigma=0.8$ also shows a
monotonically increasing density of states, but is nearly constant at low
energy. We will see that such behavior leads to a quadratic temperature
dependence of the penetration depth, even without the assumption of the
strict unitary limit. Gross \emph{et. al.} some time ago noted \cite{gross}
in a different context that T$^{2}$ behavior does not require the unitary limit.  
We note parenthetically that the behavior depicted
depends rather strongly upon the large value of impurity scattering assumed;
the first two cases will yield more exponentially activated behavior if the
scattering rate is much less strong, while the near-unitary case can
potentially \cite{PM} lead to a non-monotonic density of states.

In Figure 2 is shown the inverted squared London penetration depth $1/\lambda ^{2}(T)$,
the so-called superfluid density for several cases as indicated in the figure.
In all cases the temperature dependence of $1/\lambda ^{2}(T)$ is different
from the standard two-fluid (Gorter-Casimir) model $\lambda ^{-2}(T)=\lambda ^{-2}(0)\left[ 1-\left(
T/T_{c}\right) ^{4}\right] $, that is similar to the BCS result.
Due to the sign change between gaps,
the \textit{interband} component of the scattering matrix is strongly pair-breaking,
analogously to magnetic scattering in s-wave superconductors. As a result,
the superfluid density shows near-exponential character at low temperature in the near-Born case
($\sigma = 0.1$), while the other two cases ($\sigma = 0.4$ and $0.8$)
exhibit power-law behavior at low T, with the actual power
varying between 2 and 3.

A more detailed view of the low-temperature $\lambda(T)$ power law behavior
is presented in Figure 3, which shows $\Delta\lambda(T)/\lambda_{T=0}$ for
the same three cases. We see that the near-Born limit case ($\sigma=0.1$)
approaches a T$^{4}$ behavior, reminiscent of a two fluid model, while the
near-unitary case shows a fairly robust T$^{2}$ behavior  and the
intermediate case falls between these two limits, as one would naively
expect. Experimental data available so far \cite%
{exp_lambda1,exp_lambda2,exp_microwave} are consistent with either T$^{2}$ ,
or T%
$^{4}$ or exponential (gapped) behavior. Within our model, both results can be
explained by proper choice of the impurity scattering rate. It is
interesting to note that the $T^{2}$ dependence we obtain corresponds to strongly
gapless regime. Similar results were obtained recently in
Ref.\cite{VVC} but in the Born limit only.

\begin{figure}[tbp]
\includegraphics[width=0.8\linewidth]{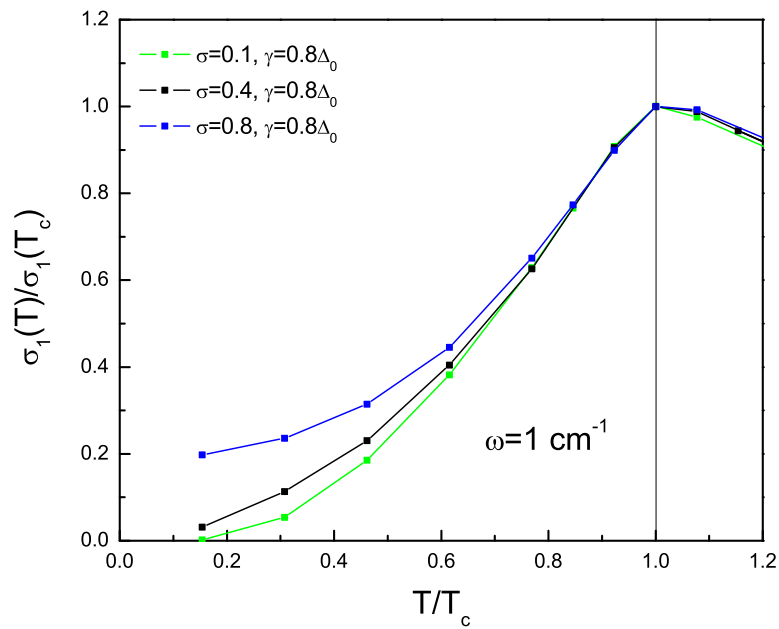}
\caption{(color online) The real part of the microwave conductivity.  Note the substantial increase with scattering strength at low temperature.}
\label{fig4}
\end{figure}

Figure 4 shows the calculated real part of the microwave conductivity for
the three cases above. The microwave conductivity( Fig. 4) $\sigma _{1}(T)$ does not
show the coherence peak near $T_{c}$. The suppression is connected with
strong-coupling effects (see, \cite{AllenRainer}). Below $T_{c}$ the
behavior of the $\sigma _{1}(T)$ is determined by the filling of the
impurity induced states below $\Delta .$ Qualitatively it is similar to the
temperature dependence of the NMR relaxation rate ( see Fig. 5), but in the
latter case the Hebel-Slichter peak is additionally reduced for $s_{\pm }$
model by the different kind of the coherence factor.
Almost all of the non-canonical BCS behavior derives from the interband
component of the scattering matrix, which results in near constant behavior at low T for the
near-unitary case, as might be expected from the form of Equation \ref
{sig_imp}, in which a squared density of states enters.
The intermediate case shows power law behavior as well, with the
precise exponent not extracted.

\begin{figure}[tbp]
\includegraphics[width=0.8\linewidth]{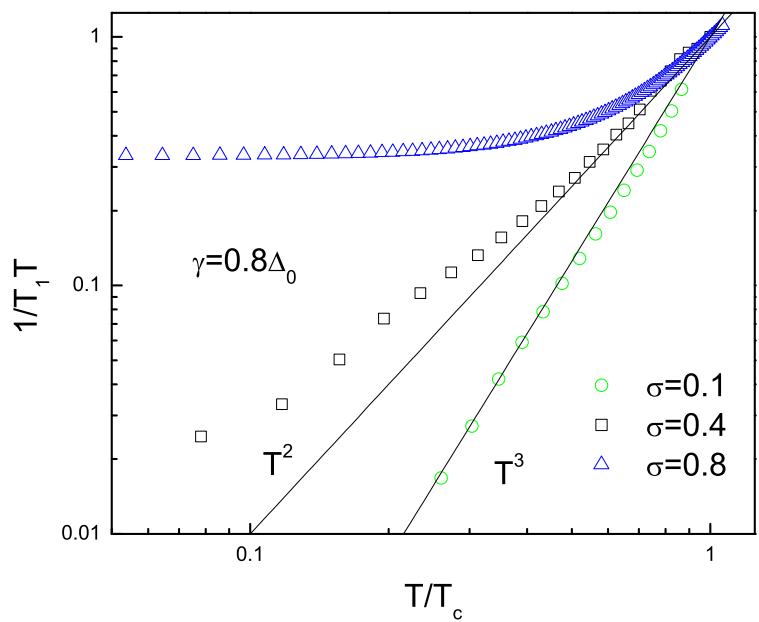}
\caption{(color online) The temperature dependence of the relaxation rate $1/T_1T$, exhibiting near-Korringa behavior for $\sigma=0.8$ and power-law behavior for the other curves.}
\label{fig4}
\end{figure}

Finally we turn in Figure 5 to the nuclear spin relaxation rate $T_{1}^{-1}$
for the same three $\sigma$ scenarios.  Note also that
following convention we have plotted $(T_{1}T)^{-1}$ rather than T$^{-1}_{1}$%
, and all power-law references here mean $(TT_{1})^{-1}$. T$_{1}$ has been a
source of substantial controversy in the pnictides due to the existence of
several data-sets \cite{NMR1,NMR2,NMR3,NMR4} showing near-T$^{2}$ behavior
throughout nearly the entire temperature range, although there now exist
data \cite{NMR5} deviating from this behavior. Several things are apparent
from the plot: first of all, the near-Born limit case shows power law
behavior ($1/T_{1}T \sim T^{3}$) throughout nearly the entire temperature
range below T$_{c}$, although it will ultimately revert to exponentially
activated behavior at the lowest temperatures. Substantial impurity scattering in the
Born limit can thus mimic much of the behavior commonly ascribed to nodes, as was noted in
\cite{Chubukov,VVC}.
The
intermediate case shows an approximate T$^{1.5}$ behavior, as was described
in \cite{Parker}, which is largely driven by the monotonic density of states
presented in Figure 1, where the same parameters are chosen. Korringa
behavior results in the near-unitary limit, as is again a direct consequence
of the corresponding behavior of the density of states in Figure 1, but does not result in either
of the first two cases unless the scattering rate $\gamma$ is increased significantly beyond
$0.8\Delta_{0}$.

It should now be clear that impurity scattering in various strengths (i.e, $%
\sigma$), if sufficient impurity concentrations are present, can produce a
wide variety of power-law behaviors in many thermodynamic quantities, even
in the near-Born limit. In the s$_{\pm}$ state, interband impurities are
clearly much more effective in creating such behavior. This has implications for the ongoing
lively debate about pairing symmetry, with significant numbers of proposals for nodal
superconductivity in the pnictides and some experimental evidence for such
behavior.

In conclusion, we have calculated the microwave response and the NMR
relaxation rate for a superconductor in $s_{\pm }$ symmetry state by solving
Eliashberg equations with a model spectrum and taking into account impurity
scattering beyond the Born limit. We show that the $T^2$ temperature
behavior of the penetration depth and the NMR relaxation rate at low
temperatures can be reproduced in this model. We have also demonstrated the
dramatic effect of the impurity scattering on the real part of the
microwave conductivity, which in particular results in near constant
behavior at low T for the near-unitary case.

\end{document}